\documentclass[12pt]{article}

\pdfoutput=1
\usepackage{epstopdf}
\usepackage{amsmath}
\usepackage{amssymb}
\usepackage{graphicx}
\usepackage{subcaption}
\usepackage{url}
\usepackage{comment}
\usepackage[sort&compress, numbers, merge]{natbib}
\usepackage[colorlinks=true, linkcolor=blue, citecolor=blue,urlcolor=black]{hyperref}

\begin{document}

\begin{titlepage}
\begin{center}

\hfill DESY 15-050 \\
\hfill IPMU 15-0046 \\
\hfill \today

\vspace{1.5cm}
\textbf{\large Indirect Probe of Electroweak-Interacting Particles\\
at Future Lepton Colliders}

\vspace{2.0cm}
\textbf{Keisuke Harigaya}$^{(a, b)}$,
\textbf{Koji Ichikawa}$^{(b)}$, 
\textbf{Anirban Kundu}$^{(c)}$, \\
\textbf{Shigeki Matsumoto}$^{(b)}$
and
\textbf{Satoshi Shirai}$^{(d)}$

\vspace{1.5cm}
\textit{
$^{(a)}${ICRR, University of Tokyo, Kashiwa, 277-8582, Japan} \\
$^{(b)}${Kavli IPMU (WPI), UTIAS, University of Tokyo, Kashiwa, 277-8583, Japan} \\
$^{(c)}${Department of Physics, University of Calcutta, Kolkata 700009, India} \\
$^{(d)}${Deutsches Elektronen-Synchrotron (DESY), 22607 Hamburg, Germany}
}

\vspace{2.0cm}
\abstract{Various types of electroweak-interacting particles, which have non-trivial charges under the $\mathrm{SU}(2)_L \times \mathrm{U}(1)_Y$ gauge symmetry, appear in various extensions of the Standard Model. These particles are good targets of future lepton colliders, such as the International Linear Collider (ILC), the Compact LInear Collider (CLIC) and the Future Circular Collider of electrons and positrons (FCC-ee). An advantage of the experiments is that, even if their beam energies are below the threshold of the production of the new particles, quantum effects of the particles can be detected through high precision measurements. We estimate the capability of future lepton colliders to probe electroweak-interacting particles through the quantum effects, with particular focus on the wino, the Higgsino and the so-called minimal dark matters, and found that a particle whose mass is greater than the beam energy by 100-1000 GeV is detectable by measuring di-fermion production cross sections with $O(0.1)$\% accuracy. In addition, with the use of the same analysis, we also discuss the sensitivity of the future colliders to model independent higher dimensional operators, and found that the cutoff scales corresponding to the operators can be probed up to a few ten TeV.}

\end{center}
\end{titlepage}
\setcounter{footnote}{0}

\section{Introduction}
\label{sec: intro}

A very wide range of models beyond the Standard Model (SM) predict extra ElectroWeak-Interacting Massive Particles (EWIMPs). One of the most promising possibilities is a weakly interacting massive particle as a dark matter candidate. For instance, a particle in a higher representation of SU(2)$_L$ attracts attention as ``Minimal Dark Matter" (MDM)\,\cite{Cirelli:2005uq, *Cirelli:2007xd, *Cirelli:2009uv}. Such a large gauge quantum number strictly restricts couplings between the particle and SM particles, and make the particle accidentally long-lived enough to be dark matter. Another concrete example is the Lightest Supersymmetric (SUSY) Particle (LSP). After the discovery of the SM-like Higgs boson with a mass around $125$\,GeV\,\cite{Aad:2012tfa, Chatrchyan:2012ufa}, the wino LSP and the Higgsino LSP attract more attention. The wino LSP is a natural prediction in the framework of the anomaly mediation\,\cite{Randall:1998uk, Giudice:1998xp} (see also\,\cite{Bagger:1999rd, *DEramo:2012qd, *Harigaya:2014sfa}) and various high-scale SUSY scenarios\,\cite{Wells:2003tf, *Wells:2004di, ArkaniHamed:2004fb, *Giudice:2004tc, *ArkaniHamed:2004yi, *ArkaniHamed:2005yv, Ibe:2006de}, which are all compatible with the 125\,GeV Higgs mass\,\cite{Okada:1990vk, *Okada:1990gg, *Ellis:1990nz, *Haber:1990aw, *Ellis:1991zd, Giudice:2011cg}. Their theoretical and phenomenological aspects are now being intensively studied\,\cite{Hall:2011jd, Hall:2012zp, Ibe:2011aa, *Ibe:2012hu, Arvanitaki:2012ps, ArkaniHamed:2012gw, Nomura:2014asa}. The Higgsino LSP is, on the other hand, predicted in the focus point like scenarios\,\cite{Feng:1999mn, *Feng:1999zg, Feng:2000gh}, which are again compatible with the observed Higgs mass while succeeding to give the electroweak scale naturally. In this article, we consider these well-motivated EWIMPs as examples.

It is essential to study the nature of such EWIMPs in order to solve the dark matter problem and understand its underlying fundamental theory. There are various ways to examine EWIMPs. If an EWIMP is a dark matter particle, its cosmic abundance and direct (indirect) signatures provide us precious information on it. If an EWIMP couples to SM particles, low-energy precise measurements of the $Z$ boson and the electric dipole moment of SM particles may reveal its nature\,\cite{Hisano:2014kua, Nagata:2014aoa}. Among various examinations, the most convincing one is a high-energy collider experiment. If an EWIMP is, however, color-neutral, hadron colliders are not powerful to search for it due to huge backgrounds, unless the EWIMP gives a very special signature like a massive charged track\,\cite{Ibe:2006de, *Buckley:2009kv, Asai:2007sw, *Asai:2008sk, *Asai:2008im, TheATLAScollaboration:2013bia, *CMS:2014gxa}. In fact, the 8\,TeV running of the Large Hadron Collider gives a poor constraint on the pure Higgsino LSP with its mass about 100\,GeV\,\cite{Han:2014kaa}, which is almost the same as the one from the 209\,GeV running of the LEP\,I\!I\,\cite{Heister:2002mn}. We therefore expect that future lepton colliders, such as the International Linear Collider (ILC), the Compact LInear Collider (CLIC) and the Future Circular Collider of $e^- e^+$ (FCC-ee), offer us the most promising way to examine EWIMPs.

An important question here is; how much beam energy is necessary to probe EWIMPs? It is obvious that a beam energy significantly above the threshold of EWIMP production is adequate to detect them utilizing e.g. the mono-photon search\,\cite{Birkedal:2004xn, Berggren:2013vfa}. However, even if the beam energy is less than the threshold, virtual EWIMP-loop corrections must contribute to SM processes. Detailed observations of the processes therefore enable us to reveal the contributions of EWIMPs. We show in this article that the SM process $e^- e^+ \to f \bar{f}$ (with $f$ being some SM fermion) is actually suitable to probe the contributions. We eventually find that an EWIMP with its mass larger than the beam energy by 100-1000\,GeV is detectable depending on its $\mathrm{SU}(2)_L \times \mathrm{U}(1)_Y$ charges when its differential cross section is measured with $O(0.1)$\% accuracy under a well-polarized beam.

The organization of this article is as follows. In Section\,\ref{sec: bsm contributions}, we consider radiative corrections to SM processes from EWIMP-loop diagrams and argue that the di-fermion production channel $e^- e^+ \to f \bar{f}$ is indeed a powerful probe. In Section\,\ref{sec: prospects}, we discuss the sensitivity of future lepton colliders to detect the EWIMP corrections with some well-motivated examples (the Wino LSP, the Higgsino LSP and some MDMs), and also study the prospect of model-independent higher dimensional operators. Section\,\ref{sec: summary} is devoted to summary of our discussion.

\section{Corrections from an EWIMP}
\label{sec: bsm contributions}

\subsection{Direct and indirect collider signatures of an EWIMP}
\label{subsec: dim6}

Let us consider an EWIMP of a mass $m\,(>0)$ which is an SU(2)$_L$ $n$-tuplet and has a hypercharge of $Y$. Here we assume that the EWIMP interacts with SM particles only through the SM gauge interactions. The EWIMP may have other renormalizable interactions with SM particles especially when it is scalar. For instance, the interaction $|\phi|^2 |H|^2$ is allowed with $\phi$ being the EWIMP. The interaction $(\phi H)^2$ or $(\phi^\dagger H)^2$ is also possible if $\phi$ has a hypercharge of $\pm 1/2$. Though these interactions contribute to the EWIMP mass after the electroweak symmetry breaking, its effect is not significant as far as the EWIMP mass $m$ is enough larger than the electroweak scale. We therefore neglect the effects in the following discussions to make our discussion simple.\footnote{The interactions addressed here are allowed even if the EWIMP is odd under some Z$_2$ symmetry while all SM particles being even, as in the case that the EWIMP is dark matter. The interactions are, however, severely constrained by recent dark matter direct detection experiments when the EWIMP plays the role of dark matter.}

Let us now start discussing physics of the EWIMP at future lepton colliders. When the center of mass energy of the collision $\sqrt{s}$ is larger than $2m$, the EWIMP can be directly pair-produced. We will then observe missing energy plus photons and/or (soft) hadrons and/or massive charged tracks accompanying with the EWIMP production\,\cite{Gunion:2001fu}.\footnote{If another new particle has a mass of $O(m)$, we may observe other collider signatures in addition to those mentioned here. See e.g. Refs.\,\cite{Ghosh:2000fv, Ghosh:2001xp} for the sequestering scenario with the anomaly mediation.} In this case, although the detail of signatures depends strongly on how the mass of each component of the EWIMP multiplet is distributed, the potential for discovery of the EWIMP in future lepton colliders is promising.

When $\sqrt{s}$ is less than $2m$, we can probe indirect effects of the EWIMP on SM processes, while the EWIMP itself cannot be directly produced. In order to study the indirect effects of the EWIMP for the case of $m \gg \sqrt{s}$, it is convenient to consider an effective field theory, namely the SM Lagrangian plus higher dimensional operators which are obtained after integrating the EWIMP out\,\cite{Henning:2014wua}. Leading contributions are given by the following dimension six operators;
\begin{align}
\Delta\mathcal{L}_{\mathrm{dim.6}} = 
-\frac{c^{\pm}_{2W}}{2\Lambda^2_{2W}} (D^\mu W^a_{\nu \rho})(D_\mu W^{a \nu \rho})
+ \frac{g\,c^{\pm}_{3W}}{6 \Lambda^2_{3W}}\varepsilon^{abc} W^{a\mu}_\rho W^{b\nu}_\mu W^{c\rho}_{\nu}
-\frac{c^{\pm}_{2B}}{2 \Lambda^2_{2B}} (\partial^\mu B_{\nu \rho}) (\partial_\mu B^{\nu \rho})\,,
\label{eq: dim6}
\end{align}
where $\Lambda$'s are suppression scales of $O(4\pi m)$, $c^\pm$’s are a sign $\pm 1$, $g$ ($g'$) is the gauge coupling of SU(2)$_L$ (U(1)$_Y$) and $W^a_{\mu\nu}$ ($B_{\mu\nu}$) is the field strength tensor of the gauge interaction SU(2)$_L$ (U(1)$_Y$), respectively, with $D_\mu$ being the covariant derivative on $W^a_{\mu\nu}$. At one-loop level,
\begin{align}
\frac{c^{\pm}_{2W}}{\Lambda^2_{2W}} &= \frac{g^2}{16\pi^2}\frac{1}{60m^2} \frac{n(n-1)(n+1)}{6}
\begin{cases}
1 & ({\rm Complex~scalar})\\
8 & ({\rm Dirac~fermion})
\end{cases}\,,\\
\frac{c^{\pm}_{3W}}{\Lambda^2_{3W}} &= \frac{g^2}{16\pi^2}\frac{1}{60m^2} \frac{n(n-1)(n+1)}{6}
\begin{cases}
1 & ({\rm Complex~scalar})\\
-2 & ({\rm Dirac~fermion})
\end{cases}\,,\\
\frac{c^{\pm}_{2B}}{ \Lambda^2_{2B}} &= \frac{g^{\prime 2}}{16\pi^2}\frac{1}{60m^2} 2nY^2
\begin{cases}
1 & ({\rm Complex~scalar})\\
8 & ({\rm Dirac~fermion})
\end{cases}\,.
\end{align}
An additional factor of $1/2$ should be multiplied for a real scalar and a Majorana fermion.

The operator involving three field strength tensors of $W^a_{\mu\nu}$ induces anomalous triple gauge couplings $\gamma WW$ and $ZWW$ (with $\gamma$, $W$ and $Z$ being photon, $W$ and $Z$ bosons), which affect e.g. the process $e^- e^+ \to W^-W^+$. For $\sqrt{s} = 1-5$\,TeV and the integrated luminosity $L = 1$\,ab$^{-1}$, it has been shown that $\Lambda_{3W} = 5-10$\,TeV can be probed through the process\,\cite{Accomando:2004sz, Baak:2013fwa}. On the other hand, as the operators involving two field strength tensors of $W^a_{\mu\nu}$ or $B_{\mu\nu}$ become four Fermi-interactions via the equations of motions of the gauge fields, the operators also affect the processes $e^- e^+ \to f \bar{f}$ (with $f$ being the SM fermion). Through these processes, the suppression scales can be probed up to $\Lambda_{2W, 2B} \sim 30 (\sqrt{s}/1\,\mathrm{TeV})^{1/2}(L/1\,\mathrm{ab}^{-1})^{1/4}$\,TeV, as we will see in the next section. We therefore expect that these di-fermion production processes will be better to probe the EWIMP indirectly.

\subsection{Corrections to di-fermion production processes}

\begin{figure}[t]
\centering
\includegraphics[clip, width = 0.47 \textwidth]{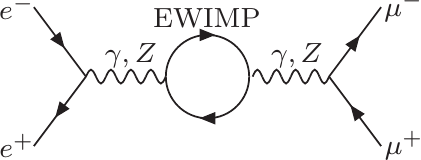}
\caption{\sl \small Corrections to the di-fermion (di-muon) production process from a fermionic EWIMP.}
\label{fig: loop}
\end{figure}

According to the argument in the previous subsection, we focus on the SM processes $e^- e^+ \to f \bar{f}$ in this article and investigate the capability of future lepton colliders to probe EWIMPs. They affect the cross sections of the processes through loop corrections even if the beam energy is smaller than $m$. An example of the corrections to the process (di-muon production process) from a fermionic EWIMP is shown in Fig.\,\ref{fig: loop}. Though we have assumed $m \gg \sqrt{s}$ in the previous subsection and used the effective field theory including dimension six operators, full form factors of the gauge boson propagators are needed for $m \gtrsim \sqrt{s}/2$. After integrating the EWIMP out at one-loop level, we obtain the following effective Lagrangian for the $e^- e^+ \to f \bar{f}$ processes:
\begin{align}
\mathcal{L}_\mathrm{eff} = \mathcal{L}_\mathrm{SM}
+ \frac{g^2 C_{WW}}{8} W^a_{\mu\nu}\,\Pi(-D^2/m^2)\,W^{a \mu \nu}
+ \frac{g^{\prime 2} C_{BB}}{8} B_{\mu\nu}\,\Pi(-\partial^2/m^2)\,B^{\mu \nu}
+ \cdots\,,
\label{eq: effective lagrangian}
\end{align}
where $\mathcal{L}_\mathrm{SM}$ stands for the SM Lagrangian and the coefficients $C_{WW}$ and $C_{BB}$ are given by
\begin{align}
C_{WW} &= \frac{n(n-1)(n+1)}{6}
\begin{cases}
1 & (\mbox{Complex scalar})\\
8 & (\mbox{Dirac fermion})
\end{cases}\,,
\label{eq: cWW}\\
C_{BB} &= 2nY^2
\begin{cases}
1 & (\mbox{Complex scalar})\\
8 & (\mbox{Dirac fermion}) 
\end{cases}\,.
\label{eq:cBB}
\end{align}
An additional factor $1/2$ should be multiplied for a real scalar and a Majorana fermion.\footnote{If the EWIMP is either a complex scalar or a Dirac fermion with $Y \neq 0$ and plays the role of dark matter, current direct detection experiments of dark matter have already ruled out this possibility, since a $Z$ boson mediated process gives a too large spin-independent scattering cross section of the EWIMP off a nucleon. These constraints can be avoided by introducing a small coupling between the Higgs and the EWIMP, which decompose the complex scalar (the Dirac fermion) into two real scalars (two Majorana fermions) after the electroweak symmetry breaking. Even a higher dimensional coupling suppressed by a scale much higher than the electroweak one is enough to achieve the mass splitting of $\gtrsim O(100)$\,keV and thus to avoid the constraints. This is in fact the case of the Higgsino dark matter in the high-scale SUSY scenario\,\cite{Nagata:2014wma}. Such a highly suppressed coupling does not alter our discussion at all.} The ellipsis at the end of the Lagrangian includes operators composed of the strength tensors more than two, but those are irrelevant for the following discussion. The function $\Pi(x)$ is the renormalized self-energy of the gauge bosons from the EWIMP's loop. Its explicit form is
\begin{align}
\Pi(x) = 
\begin{cases}
\dfrac{1}{16\pi^2} \displaystyle\int_0^1 dy\,y (1 - y) \ln [1 - y(1-y)\,x] & (\mbox{Fermion})
\\
\dfrac{1}{16\pi^2} \displaystyle\int_0^1 dy\,(1 - 2y)^2 \ln [1 - y(1-y)\,x] & (\mbox{Scalar})
\end{cases}\,.
\end{align}
Here we have used the $\overline{\mathrm{MS}}$ regularization scheme with the renormalization scale of $\mu = m$.

As can be seen in the effective Lagrangian\,\eqref{eq: effective lagrangian}, the EWIMP's effect is encoded in the operators involving two field strength tensors, as expected from the argument in the previous subsection. This effect is highly suppressed when the typical energy scale of processes under consideration is much smaller than the EWIMP mass $m$. It is worth notifying that the absence of couplings to any SM particles other than the gauge interactions ensure the SM symmetry like flavor, CP and custodial symmetries, etc. Precision measurements at low energy experiments are thus not efficient to see the effect of the EWIMP. \footnote{
Contribution to the oblique parameters\,\cite{Peskin:1991sw} (including the extention of S, T, U\,\cite{Maksymyk:1993zm, *Barbieri:2004qk}) from the operators proportional to $C_{WW}$ and $C_{BB}$ has been evaluated in Ref.\,\cite{Martin:2004id} and turned out to be small.
}
Energetic lepton colliders such as the ILC, the CLIC and the FCC-ee will therefore play an important role to detect the new EWIMP.

Let us calculate the indirect corrections to the process $e^- e^+ \to f \bar{f}$. When the final state is a SM fermion pair other than an electron-positron pair, the leading order (LO) amplitude is
\begin{align}
i\mathcal{M}_\mathrm{LO}[e^-_h(p) e^+_{\bar{h}}(p') \to f_{h'}(k) \bar{f}_{\bar{h}'}(k')] = 
i[\bar{v}_h(p') \gamma^\mu u_h(p)]
\sum_{V = \gamma,\,Z}
\frac{C_{e_hV}\,C_{f_{h'}V}}{s - m^2_V }
[\bar{u}_{h'}(k) \gamma_\mu v_{h'}(k')]\,,
\label{eq: LO1}
\end{align}
where $h, h' = L, R$ $(\bar{h}, \bar{h}' = R, L)$ represent the chirality of the fermions. Fermion wave functions are defined as $u_{L(R)}(p) = P_{L(R)} u(p)$ and $v_{L(R)}(p) = P_{L(R)} v(p)$ with $u(p)$ and $v(p)$ being those of particles and anti-particles. Gauge couplings of the fermions are given by $C_{f_L Z} = g_Z (T_{3f}/2 - Q_f s_W^2)$, $C_{f_R Z} = -g_Z Q_f s_W^2$ and $C_{f_L\,\gamma} = C_{f_R\,\gamma} = e Q_f$ with $s_W$, $T_{3f}$ and $Q_f$ being the sine of the Weinberg angle, the third component of the weak isospin and the electric charge of the fermion $f$. The coupling $g_Z$ is defined by $g_Z = g/c_W$ with $c_W$ being the cosine of the Weinberg angle.

When the final state is an electron-positron pair, the LO amplitude of processes $e^-_L e^+_R \to e^-_R e^+_L$ and $e^-_R e^+_L \to e^-_L e^+_R$ are again given by the amplitude\,\eqref{eq: LO1}. On the other hand, those of other processes, $e^-_L e^+_R \to e^-_L e^+_R$, $e^-_R e^+_L \to e^-_R e^+_L$, $e^-_L e^+_L \to e^-_L e^+_L$ and $e^-_R e^+_R \to e^-_R e^+_R$, are given by
\begin{align}
i\mathcal{M}_\mathrm{LO}(e^-_h e^+_{\bar{h}'} \to e^-_h e^+_{\bar{h}'})
& = i[\bar{v}_{h'}(p') \gamma^\mu {u}_h(p)]
\sum_{V = \gamma,\,Z}
\frac{C_{e_hV}\,C_{e_{h'}V}}{s - m^2_V }
[\bar{u}_h(k) \gamma_\mu v_{h'}(k')] 
\nonumber \\
& - i[\bar{u}_h(k) \gamma^\mu {u}_h(p)]
\sum_{V = \gamma,\,Z}\
\frac{C_{e_hV}\,C_{e_{h'}V}}{t - m^2_V }
[\bar{v}_{h'}(p') \gamma_\mu v_{h'}(k')]\,.
\label{eq: LO2}
\end{align}

The contribution of the EWIMP to the di-fermion processes appear at the next leading order (NLO), which can be evaluated by the effective Lagrangian\,\eqref{eq: effective lagrangian}. When the final state is a SM fermion pair other than an electron-positron pair, the contribution to the amplitude is given by
\begin{align}
i\mathcal{M}_\mathrm{BSM}(e^-_h e^+_{\bar{h}} \to f_{h'} \bar{f}_{\bar{h}'}) = 
i[\bar{v}_{h}(p') \gamma^\mu u_h(p)]
\sum_{V\,V' = \gamma,\,Z}
\frac{C_{e_hV}\,C_{f_{h'}V'}\,d_{VV'}\,s\,\Pi(s)\,
[\bar{u}_{h'}(k) \gamma_\mu v_{h'}(k')]}
{(s - m^2_V ) (s - m^2_{V'})}\,,
\label{eq: NLO1}
\end{align}
where $d_{VV'}$ are gauge group factors whose explicit forms are $d_{ZZ} = (g_Z^2/2)(c_W^4 C_{WW} + s_W^4 C_{BB})$, $d_{\gamma\gamma} = (e^2/2)(C_{WW} + C_{BB})$ and $d_{Z\gamma} = d_{\gamma Z} = (e\,g_Z/2) (c_W^2 C_{WW} - s_W^2 C_{BB})$, respectively. The EWIMP's contribution to processes $e^-_L e^+_R \to e^-_R e^+_L$ and $e^-_R e^+_L \to e^-_L e^+_R$ are also given by the above formula. The contribution to the process $e^-_h e^+_{\bar{h}'} \to e^-_h e^+_{\bar{h}'}$ is, on the other hand, given by
\begin{align}
i\mathcal{M}_\mathrm{BSM}(e^-_h e^+_{\bar{h}'} \to e_h \bar{e}_{\bar{h}'})
&= i[\bar{v}_{h'}(p') \gamma^\mu u_h(p)]
\sum_{V\,V' = \gamma,\,Z}
\frac{C_{e_hV}\,C_{e_{h'}V'}\,d_{VV'}\,s\,\Pi(s)\,
[\bar{u}_h(k) \gamma_\mu v_{h'}(k')]}
{(s - m^2_V) (s - m^2_{V'})}
\nonumber \\
&- i[\bar{u}_h(k) \gamma^\mu u_h(p)]
\sum_{V\,V' = \gamma,\,Z}
\frac{C_{e_hV}\,C_{e_{h'}V'}\,d_{VV'}\,t\,\Pi(t)\,
[\bar{v}_{h'}(p') \gamma_\mu v_{h'}(k')]}
{(t - m^2_V) (t - m^2_{V'})}\,.
\label{eq: NLO2}
\end{align}
The NLO amplitudes \eqref{eq: NLO1} and \eqref{eq: NLO2} have the same chirality structure as the LO ones \eqref{eq: LO1} and \eqref{eq: LO2}. The dominant contribution of the EWIMP is thus from interference between these amplitudes.

\begin{figure}[t]
\includegraphics[width=0.48\textwidth]{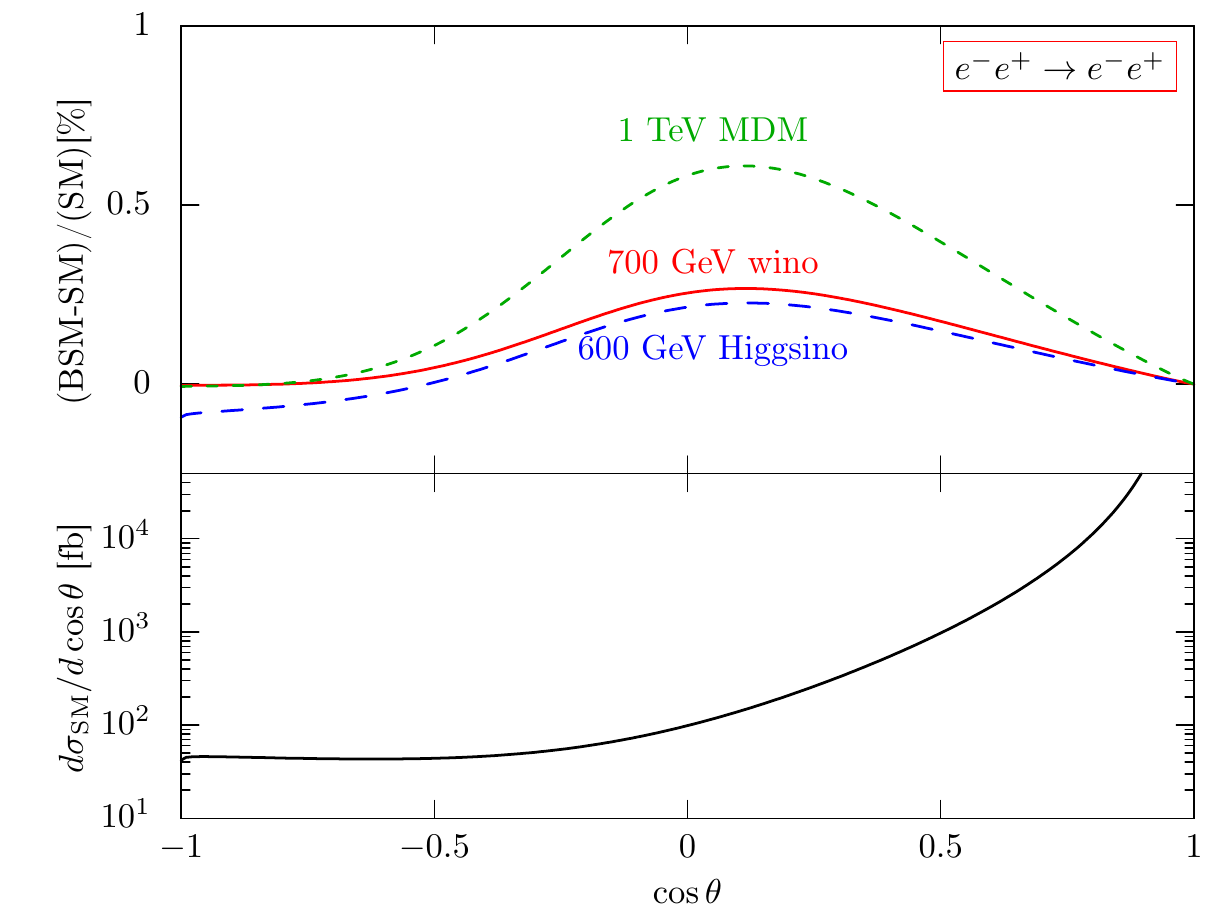}
\includegraphics[width=0.48\textwidth]{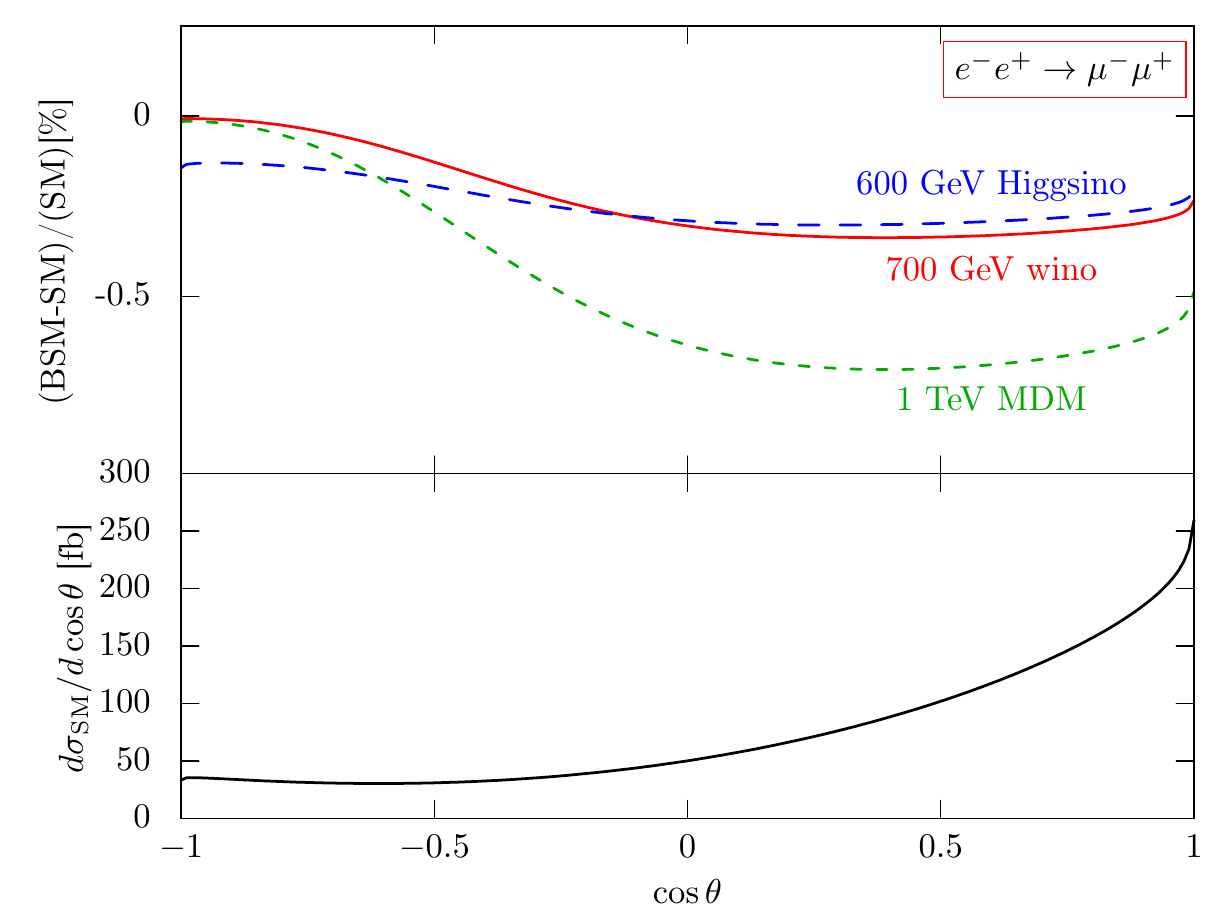}
\\
\vspace{0.5cm}
\includegraphics[width=0.48\textwidth]{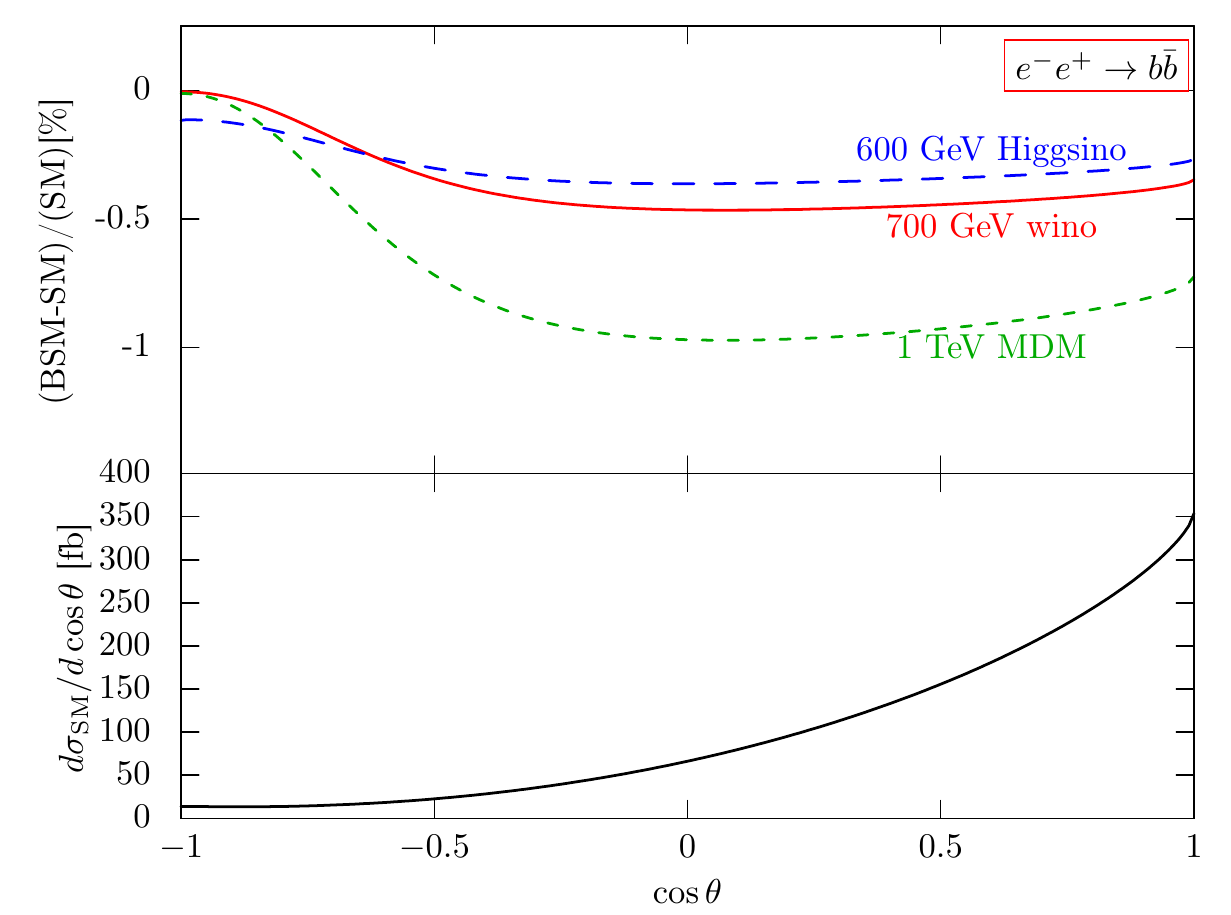}
\includegraphics[width=0.48\textwidth]{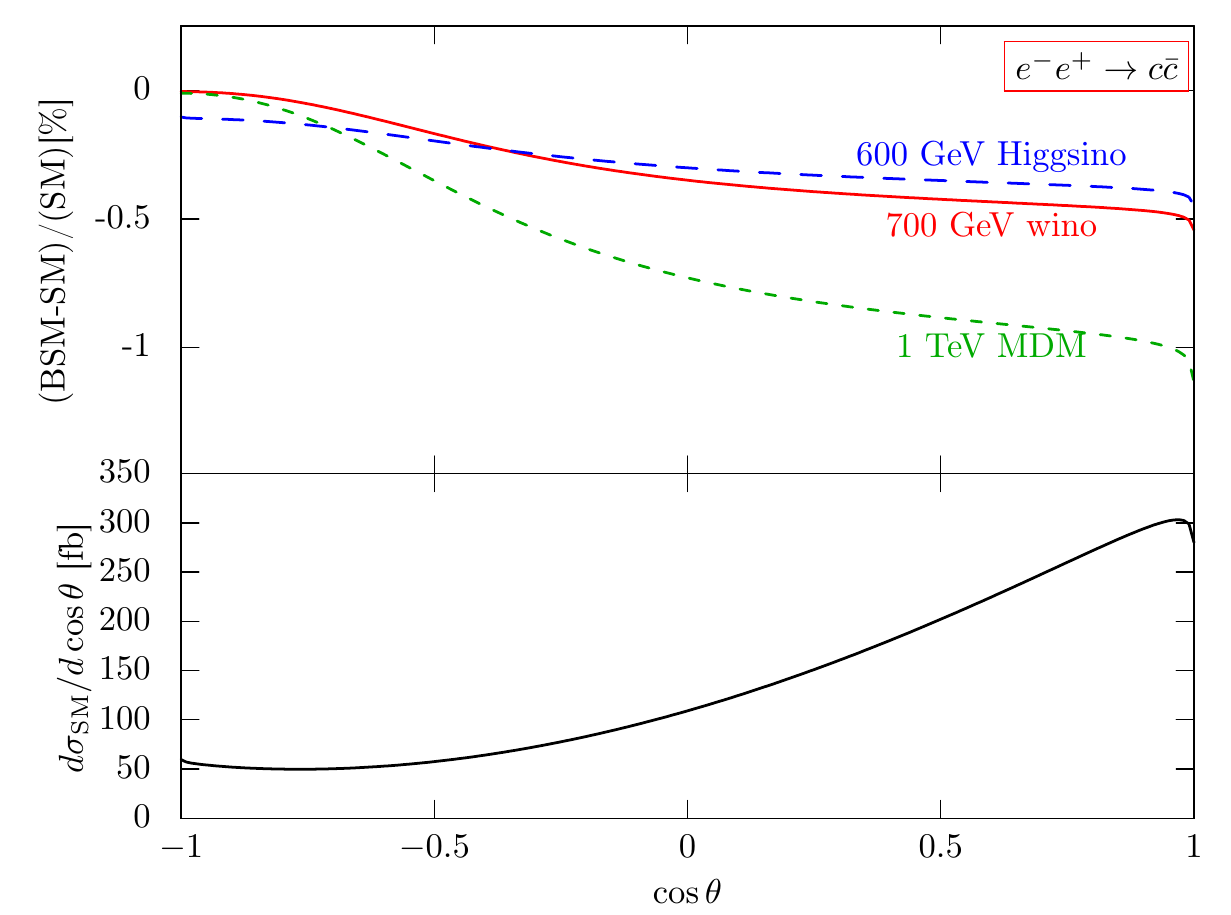}
\caption{\sl \small Contributions to the differential cross section of $e^- e^+ \to f \bar{f} (\gamma)$ from the 700\,GeV wino, the 600\,GeV Higgsino and the 1\,TeV fermionic minimal dark matter. The center of mass energy is fixed to be $\sqrt{s} = 1$\,TeV with polarizations of incoming electron and positron beams being $P_- = -80\%$ and $P_+ = 60\%$, respectively. The ``$d\sigma_\mathrm{SM}/d \cos \theta$" plots show the one-loop SM differential cross sections.}
\label{fig: xsections}
\end{figure}

As some examples, we show in Fig.\,\ref{fig: xsections} contributions to the differential cross section of $e^- e^+ \to f \bar{f} (\gamma)$ from the 700\,GeV wino (Majorana fermion with $n = 3$ \& $Y = 0$), the 600\,GeV Higgsino (Dirac fermion with $n = 2$ \& $Y = \pm 1/2$) and the 1\,TeV minimal DM (Majorana fermion with $n = 5$ \& $Y = 0$) at one-loop level with the center of mass energy of $\sqrt{s} = 1$\,TeV. Polarizations of incoming electron and positron beams are assumed to be $P_-=-80\%$ (left-handed like) and $P_+ = 60\%$ (right-handed like), respectively,\footnote{The polarization of the positron beam is assumed to be the future upgradeable maximum at the ILC\,\cite{Behnke:2013xla, *Baer:2013cma, *Adolphsen:2013jya, *Adolphsen:2013kya, *Behnke:2013lya}.} because the EWIMPs in the examples mainly affect the SU(2)$_L$ gauge boson propagator through the interaction $W^a_{\mu\nu}\Pi(-D^2/m^2)W^{a \mu \nu}$ in the effective Lagrangian \eqref{eq: effective lagrangian}. While the most important benefit of the beam polarization is to enhance the cross section and effectively increase the integrated luminosity, it also reduces the right-handed electron process which in practice contributes to the background. Therefore, the polarization can enhance the signal significance against the systematic errors, which is an additional gain to the increase of the effective luminosity. In order to depict the figures (and to discuss the prospect of future lepton colliders in the next section), we have also included SM contributions at NLO order using the code \textsc{aITALC}\,\cite{Lorca:2004fg} with a slight modification, which integrates the programs \textsc{Qgraf}\,\cite{Nogueira:1991ex}, \textsc{Diana}\,\cite{Tentyukov:1999is}, \textsc{Form}\,\cite{Vermaseren:2000nd}, \textsc{LoopTools}\,\cite{Hahn:1998yk} and \textsc{FF}\,\cite{vanOldenborgh:1990yc}. Here we set $E^{\max}_{\gamma} = 0.1 \sqrt{s}$ for $e^- e^+ \to f \bar{f} \gamma$ at the NLO calculation.

\section{Prospects of indirect signatures of EWIMPs}
\label{sec: prospects}

\subsection{Analysis method and detector performance}
\label{subsec: analysis}

In order to quantitatively investigate the capability of future lepton colliders for probing the EWIMP, we adopt the binned likelihood analysis on the differential cross section of the process $e^- e^+ \to f \bar{f}$. We use ten uniform intervals for the scattering angle $\cos\theta \in [-1:1]$ for the final state $f\ne e^-$, while $\cos\theta \in [-0.99:0.99]$ for $f=e^-$. We have assumed a simplified setup for detection efficiency; 100\% for leptons, 80\% for $b$-jets and 50\% for $c$-jets. Here, we require at least one heavy flavor quark identification for $b$- and $c$-jets channels. We then define the $\chi^2$ function as
\begin{align}
\chi^2 = \sum_{i = 1}^{10} 
\frac{\left[N_i^\mathrm{(BSM + SM)} - N_i^\mathrm{(SM)}\right]^2}
{ N_i^\mathrm{(SM)} + \left[\epsilon_i\,N_i^\mathrm{(SM)}\right]^2}\,,
\label{eq: chi2}
\end{align}
where $N_i^\mathrm{(SM + BSM)}$ ($N_i^\mathrm{(SM)}$) is the expected value of the number of events with (without) the EWIMP contribution, while $\epsilon_i$ represents a systematic error in the estimation of $N_i^\mathrm{(SM)}$. The denominator thus represents a quadratic sum of the systematic and statistical errors. We have also assumed that the correlation between the errors is negligible and treatedF them as independent ones.

We have only considered the irreducible background from the SM di-fermion process to estimate $N_i^\mathrm{(SM)}$, as discussed in the previous section. Other reducible backgrounds are expected to be negligible, for those give little events in the signal region of $E_f + E_{\bar{f}} \simeq \sqrt{s}$. In reality, the estimation of the irreducible background suffers from various kinds of experimental uncertainties, such as luminosity, polarization and acceptance estimation errors. The above $\epsilon_i$ represents a collective parameterization of these uncertainties, which is expected to be $O(0.1 - 1)\%$ according to the current ILC technical design report (TDR)\,\cite{Behnke:2013xla, *Baer:2013cma, *Adolphsen:2013jya, *Adolphsen:2013kya, *Behnke:2013lya}. Estimating the precise value of the $\epsilon_i$ is beyond the scope of this article. Instead, we examine how large the change of $\epsilon_i$ alters the capability of future lepton colliders by adopting several representative values of $\epsilon_i = 0$, 0.1, 0.3, 0.5 and 1\%.

The deviation from the SM prediction approximately scales as $N^\mathrm{(BSM + SM)} - N^\mathrm{(SM)} \propto 1/m^2$, because it comes from a interference between the SM and the EWIMP amplitudes. According to the $\chi^2$ function\,\eqref{eq: chi2}, when the statistical error dominates the systematic one, the EWIMP mass reach turns out to be approximately proportional to $s^{1/4} {L}^{1/4}$ with $L$ being the integrated luminosity. It can therefore be seen that increasing the luminosity $L$ is equivalent to increasing the collision energy squared $s$, so that accumulating data at future lepton colliders has a great impact on the EWIMP search. Needless to say, the systematic error eventually dominates the statistical one when $L$ becomes large enough, which is estimated to be $L \gtrsim (10^{-5}/\epsilon_i^2) (s/1\,{\rm TeV}^2)[{\rm ab}^{-1}]$. The mass reach is then proportional only to $s$ in such a case.

\subsection{Examples}
\label{sec: examples}

We are now at the position to discuss the capability of future lepton colliders to probe EWIMPs. As mentioned earlier, we consider several well-motivated EWIMP candidates: the wino LSP, the Higgsino LSP and a few minimal dark matters (MDMs). For the sake of convenience, we also consider the capability in terms of higher-dimensional operators discussed in section \ref{subsec: dim6} in order to provide model-independent perspectives. We have assumed the integrated luminosity of $L =3~\mathrm{ab}^{-1}$ and the beam polarizations of $P_- = -80$\% and $P_+ = +60$\% in the following discussions, unless otherwise stated explicitly.

\begin{figure}[p!]
\centering
\subcaptionbox{The wino LSP \label{fig: wino_pros}}
{\includegraphics[clip, width = 0.98 \textwidth]{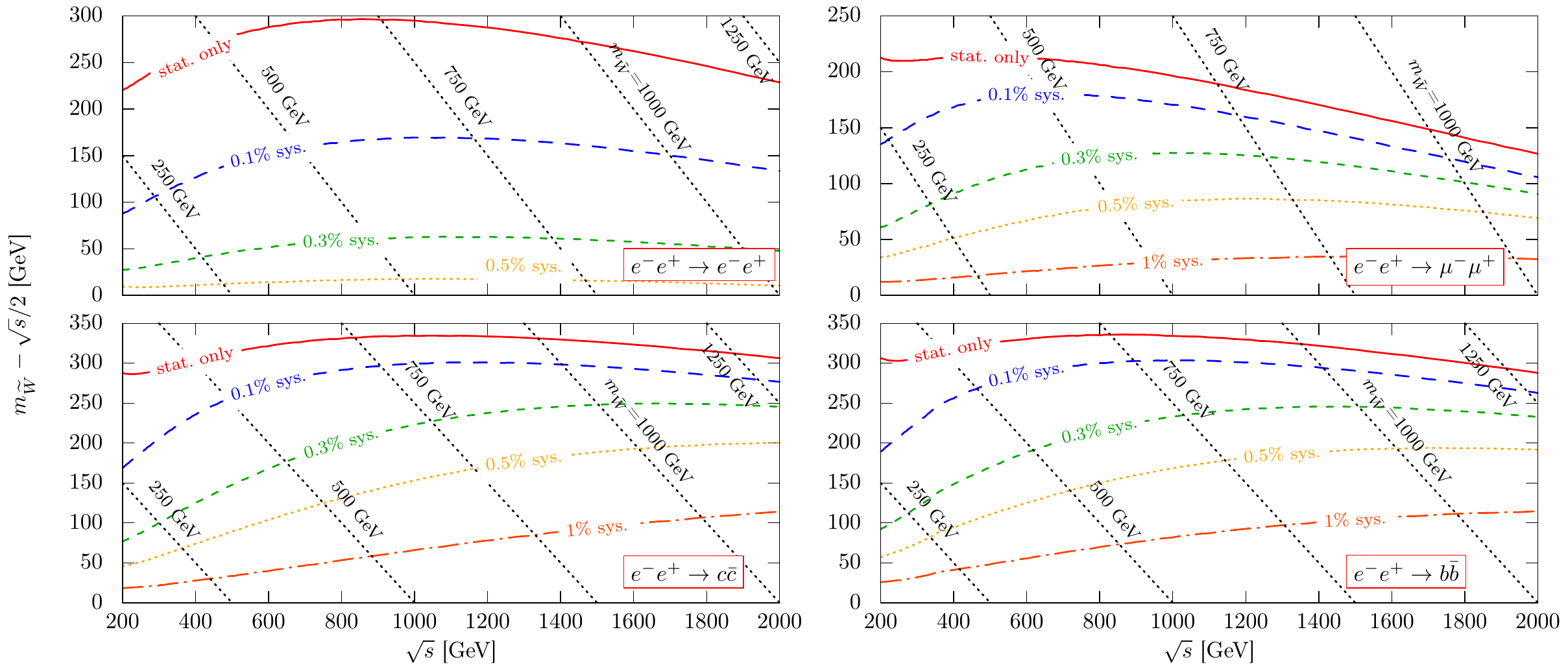}}
\subcaptionbox{The Higgsino LSP \label{fig: higgsino_prps}}
{\includegraphics[clip, width = 0.98 \textwidth]{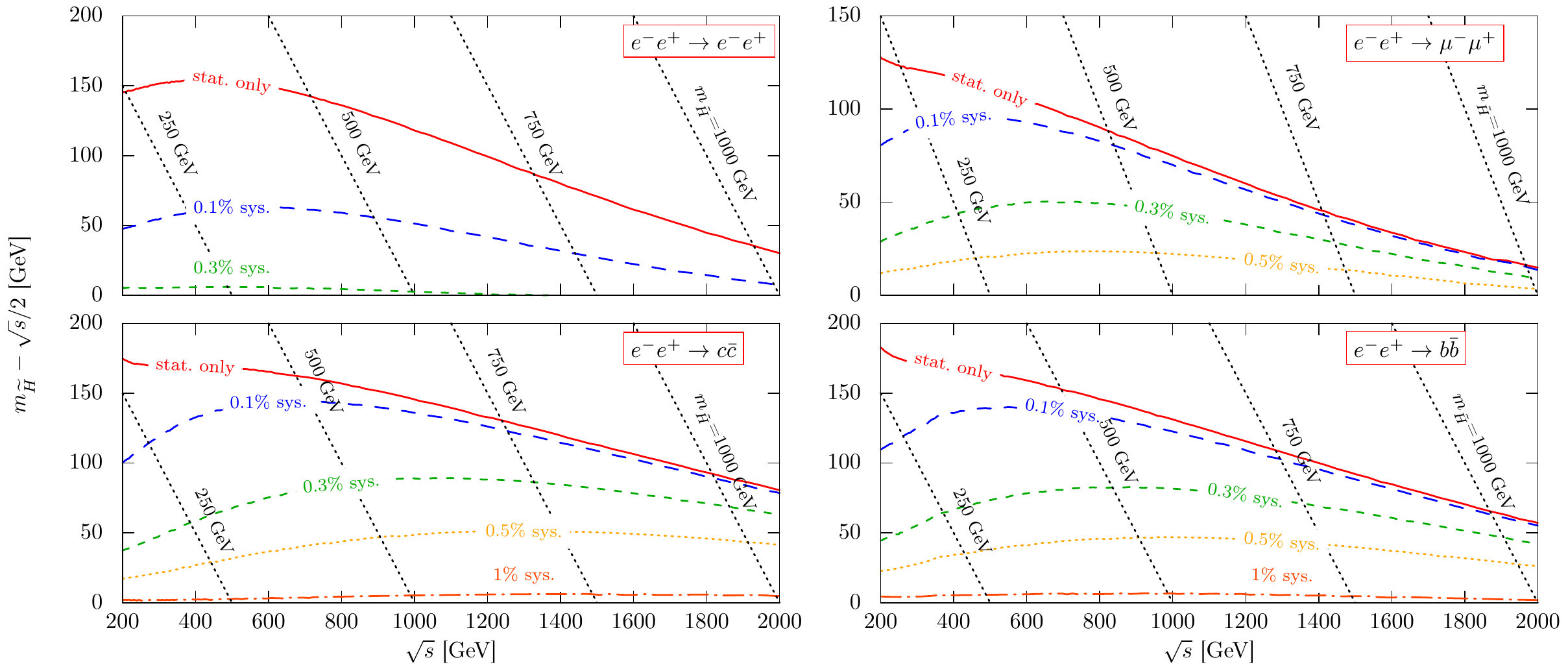}}
\caption{\sl \small Prospect of LSP dark matter searches: (a) The wino LSP ($n=3$ \& $Y=0$, Majorana fermion) and (b) The Higgsino ($n=2$ \& $Y=\pm1/2$, Dirac fermion). The differences between the expected reach of the EWIMP mass $m$ at 95\% C.L. and the beam energy $\sqrt{s}/2$ is shown. Here we assume that the integrated luminosity of $3\,\mathrm{ab}^{-1}$ and the electron and positron beam polarizations of $-80\%$ and $60\%$. We have shown the results with the systematic uncertainty of $\epsilon_i = 0$, $0.1$, $0.3$, $0.5$ and $1$\%.}
\label{fig: susy prospect}
\end{figure}

\begin{figure}[p!]
\centering
\subcaptionbox{The minimal fermion dark matter ($n=5$ and $Y=0$) \label{fig:mdm_f_pros}}
{\includegraphics[clip, width = 0.98 \textwidth]{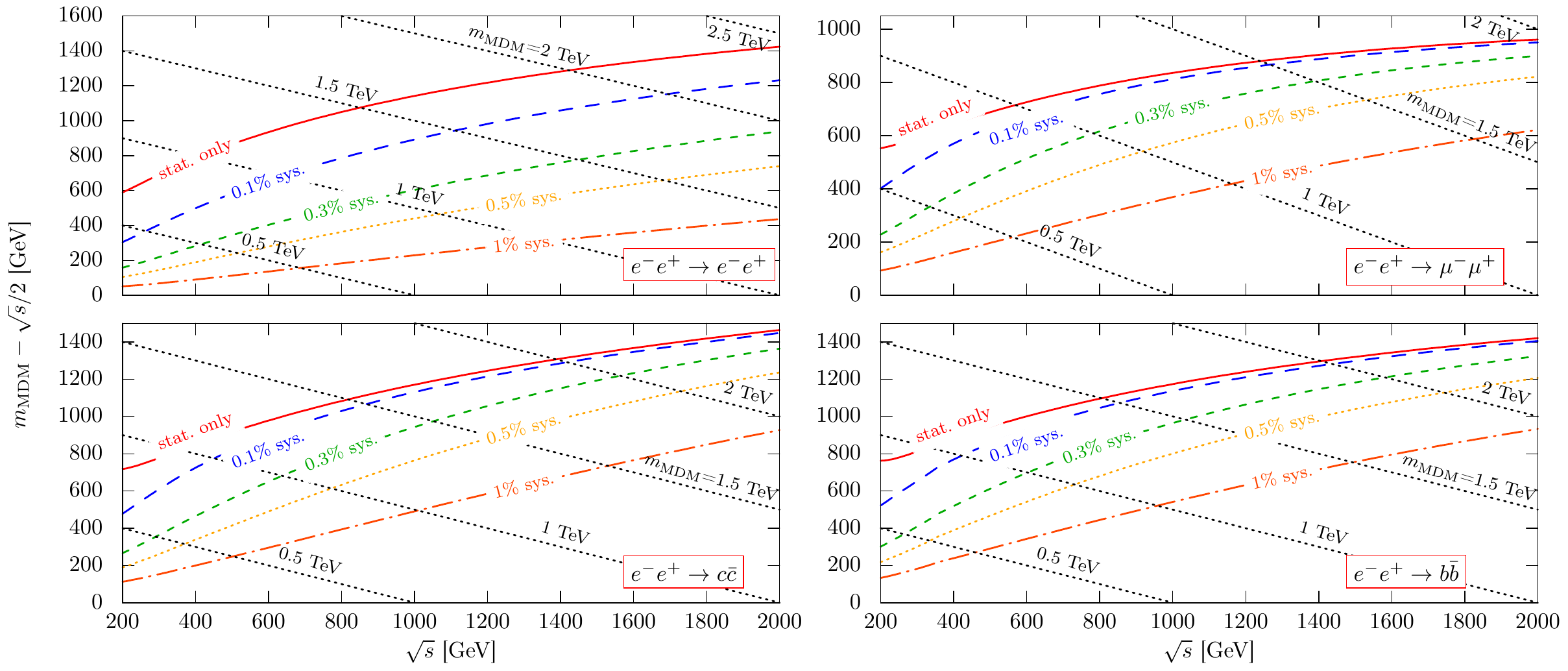}}
\subcaptionbox{The minimal scalar dark matter ($n=7$ and $Y=0$) \label{fig:mdm_s_pros}}
{\includegraphics[clip, width = 0.98 \textwidth]{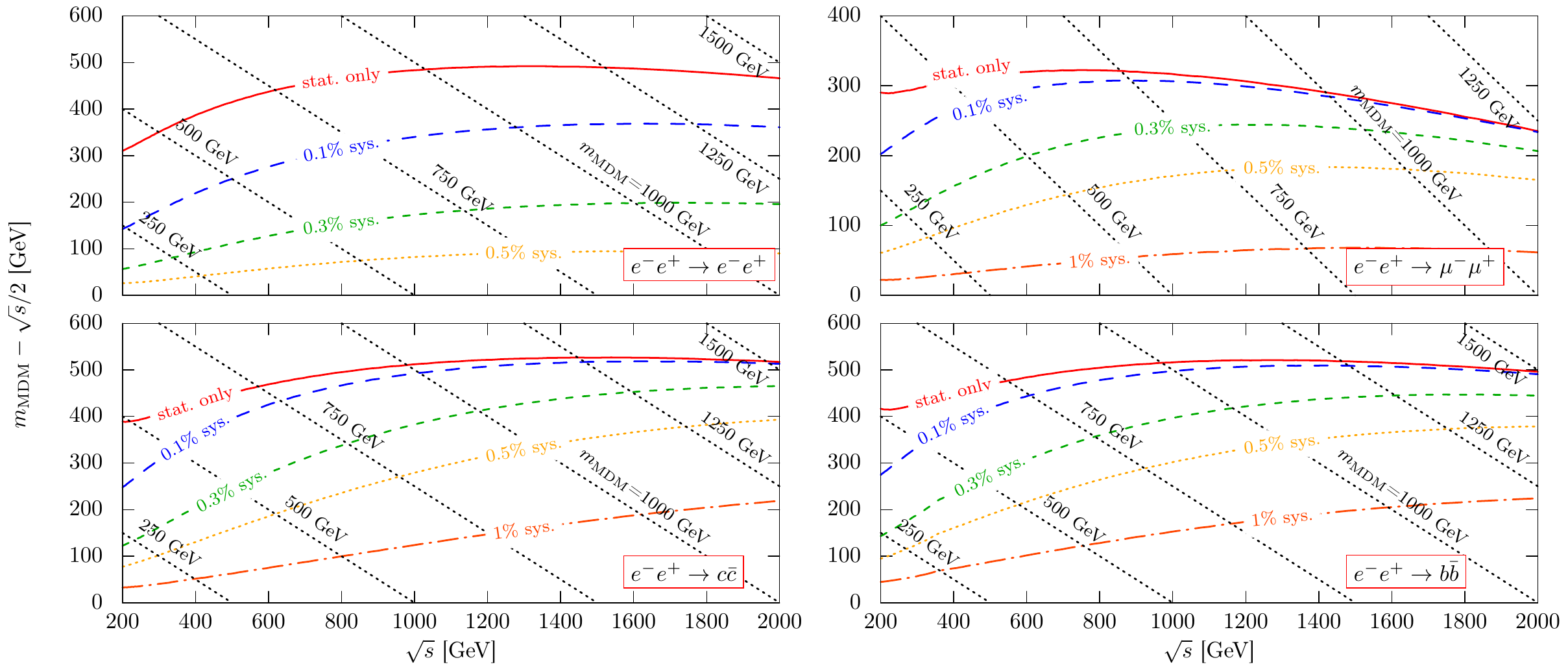}}
\caption{\sl \small Prospect of minimal dark matter (MDM) searches: (a) The Majorana fermion MDM ($n=5$ \& $Y=0$) and The real scalar MDM ($n=7$ \& $Y=0$). The differences between the expected reach of the MDM mass $m_\mathrm{MDM}$ at 95\% C.L. and the beam energy $\sqrt{s}/2$ is shown. Here we assume that the integrated luminosity of $3\,\mathrm{ab}^{-1}$ and the electron and positron beam polarizations of $-80\%$ and $60\%$. We have shown the results with the systematic uncertainty of $\epsilon_i = 0$, $0.1$, $0.3$, $0.5$ and $1$\%.}
\label{fig: mdm prospect}
\end{figure}

\begin{figure}[p!]
\centering
\subcaptionbox{The operator $(D^\mu W^a_{\mu \nu})^2$ \label{fig:WW_pros}}
{\includegraphics[clip, width = 0.98 \textwidth]{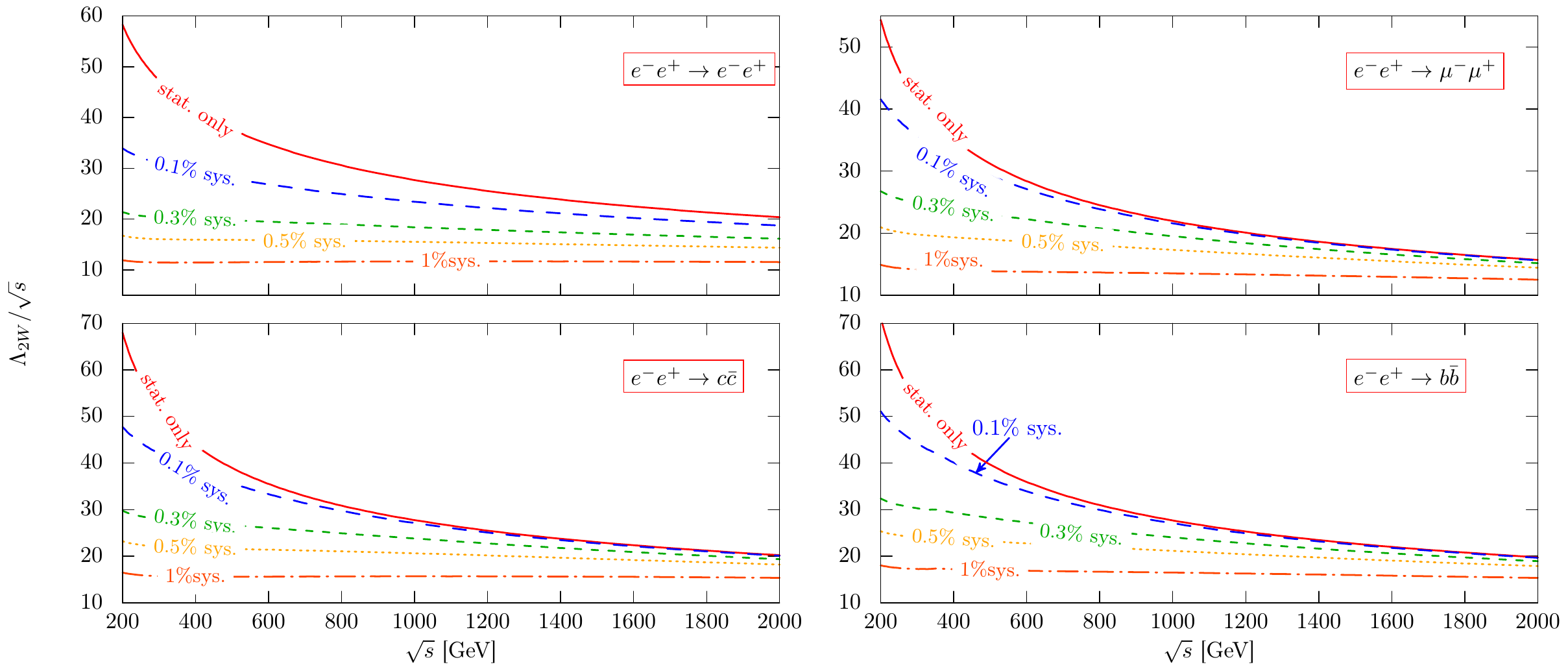}}
\subcaptionbox{The operator $(\partial^\mu B_{\mu \nu})^2 $\label{fig:BB_pros}}
{\includegraphics[clip, width = 0.98 \textwidth]{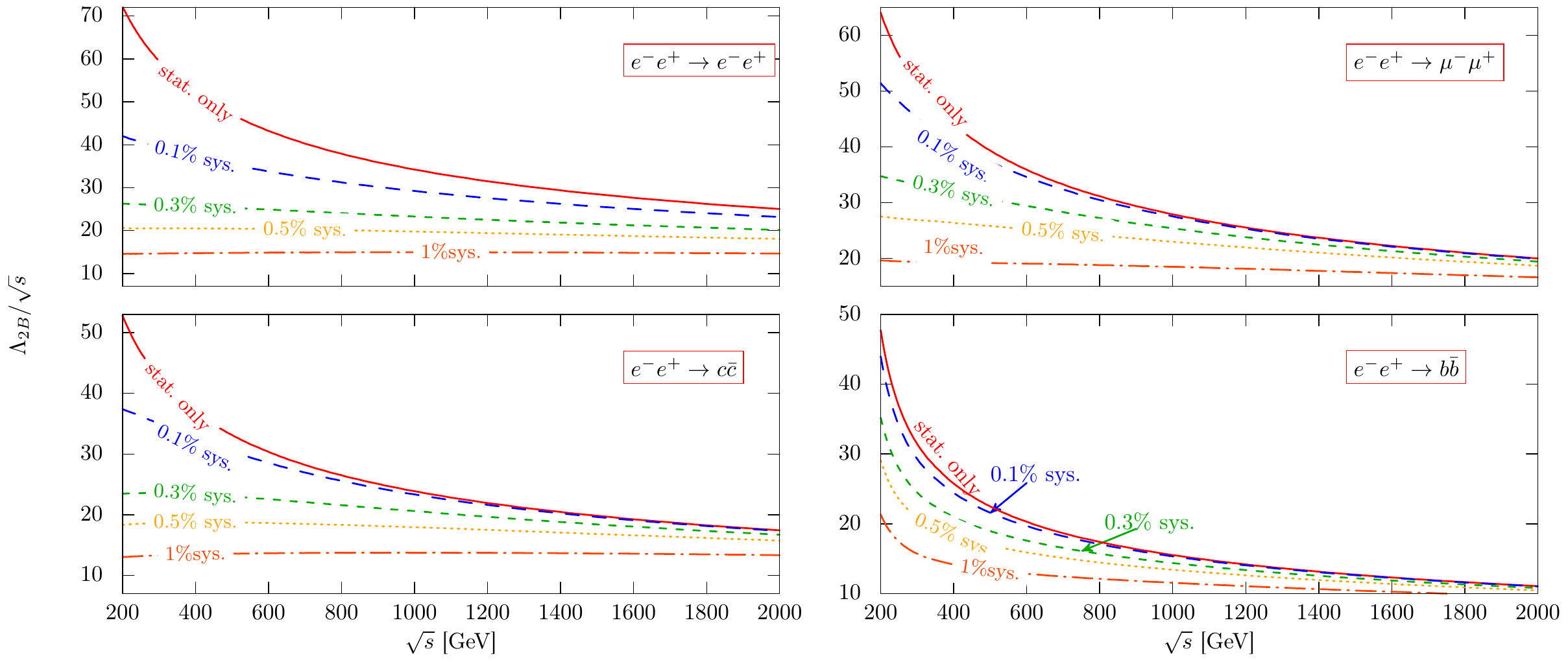}}
\caption{\sl \small Prospect of dimension six operator searches: (a) The operator $(D^\mu W^a_{\mu \nu})^2$ and (b) $(\partial^\mu B_{\mu \nu})^2$. The ratio between the expected reach of the cut-off scale $\Lambda$ at 95\% C.L. and the center of mass energy $\sqrt{s}$ is shown. Here we assume the integrated luminosity of $3~\mathrm{ab}^{-1}$ and the electron and positron beam polarizations of $-80\%$ and $60\%$ for the case (a) while $P_-=+80$\% and $P_+=-60\%$ for the case (b). We have shown the results with the systematic uncertainty of $\epsilon_i = 0$, $0.1$, $0.3$, $0.5$ and $1$\%.}
\label{fig: higher_prospect}
\end{figure}

\subsubsection*{LSP dark matters}

The capability of future lepton colliders to probe the wino LSP ($n=3$ \& $Y=0$, Majorana fermion) and Higgsino LSP ($n=2$ \& $Y=\pm1/2$, Dirac fermion) are shown in Fig.\,\ref{fig: susy prospect}. As we have mentioned before, left-handed electron and right-handed positron beam polarizations have better sensitivity than the opposite ones for these EWIMP candidates. The Higgsino LSP also affects the $\mathrm{U}(1)_Y$ gauge boson propagator, but this contribution is less significant than that from the $\mathrm{SU}(2)_L$ gauge boson propagator due to the smallness of the $\mathrm{U}(1)_Y$ gauge coupling. As can be seen from the figure, a future collider with $\sqrt{s} = 500$\,GeV (1\,TeV) will probe the mass up to 430\,GeV (670\,GeV) for the wino LSP and 340\,GeV (560\,GeV) for the Higgsino LSP by measuring the scattering cross section of the $e^-e^+ \to \mu^-\mu^+$ process with the systematic error of $\epsilon_i = 0.1$\%.

\subsubsection*{Minimal dark matters}

We consider two types of the MDM. One is a Majorana fermion with $n=5$ and $Y=0$ and the other is a real scalar with $n=7$ and $Y=0$. The stability of both particles is automatically guaranteed without imposing any ad hoc parities\,\cite{Cirelli:2005uq, *Cirelli:2007xd, *Cirelli:2009uv}. The capability of future lepton colliders to probe these dark matters are shown in Fig.\,\ref{fig: mdm prospect}. Left-handed electron and right-handed positron beam polarizations are better than the opposite ones in both cases. As can be seen from the figure, a future collider with $\sqrt{s} = 500$\,GeV (1\,TeV) will probe the mass up to 850\,GeV (1.5\,GeV) for the fermionic MDM and 530\,GeV (810\,GeV) for the bosonic MDM by measuring the scattering cross section of the $e^-e^+ \to \mu^-\mu^+$ process with the systematic error of $\epsilon_i = 0.1$\%. The contribution of the EWIMP to di-fermion processes is approximately proportional to $n^3$ as seen in Eq.\,\eqref{eq: cWW}, so that the minimal dark matters are more easily accessible than the LSP dark matters.

\subsubsection*{Dimension-six operators}

Here we consider the discovery reach of the higher dimensional operators $(D^\mu W^a_{\mu \nu})^2$ and $(\partial^\mu B_{\mu \nu})^2 $ in Eq.\,\eqref{eq: dim6}. Signs $c^{\pm}_{2W}$ and $c^{\pm}_{2B}$ are fixed to be one, while other choices of the signs does not alter the result much. Expected limits on the cut-off scales $\Lambda_{2W}$ and $\Lambda_{2B}$ are shown in Fig.\,\ref{fig: higher_prospect}. Left-handed electron and right-handed positron beam polarizations provide better sensitivity for $(D^\mu W^a_{\mu \nu})^2$, while right-handed electron and left-handed positron beam polarizations are better for $(\partial^\mu B_{\mu \nu})^2$, for the right-handed electron has a larger $\mathrm{U}(1)_Y$ gauge charge than that of the left-handed one. We therefore assume the electron and positron beam polarizations of $80\%$ and $-60\%$ for the $(\partial^\mu B_{\mu \nu})^2$ case. It can be seen from the figure that a future collider with $\sqrt{s} = 500$\,GeV (1\,TeV) will reach the scale of 15\,TeV (22\,TeV) for $\Lambda_{2W}$ and 19\,TeV (28\,TeV) for $\Lambda_{2B}$ by measuring the scattering cross section of the $e^-e^+ \to \mu^-\mu^+$ process with the systematic error of $\epsilon_i = 0.1$\%.

\subsection{Potential of future lepton colliders with large $\sqrt{s}$}

We have discussed the setup motivated mainly by the proposed TDR of the ILC project so far. It would be also interesting to investigate how heavy dark matter can be in principle probed at future lepton colliders with very high energy center of mass energy. In such colliders with beam energy much higher than the TeV scale, the statistical error tends to dominate the systematic one, for di-fermion production cross sections scales as $1/s$. We therefore neglect the systematic uncertainty in this investigation and combine the $e$, $\mu$, $c$ and $b$ channels in the analysis in order to estimate the ultimate potential of future lepton colliders for the EWIMP search.

\begin{figure}[t]
\centering
\includegraphics[clip, width = 0.98 \textwidth]{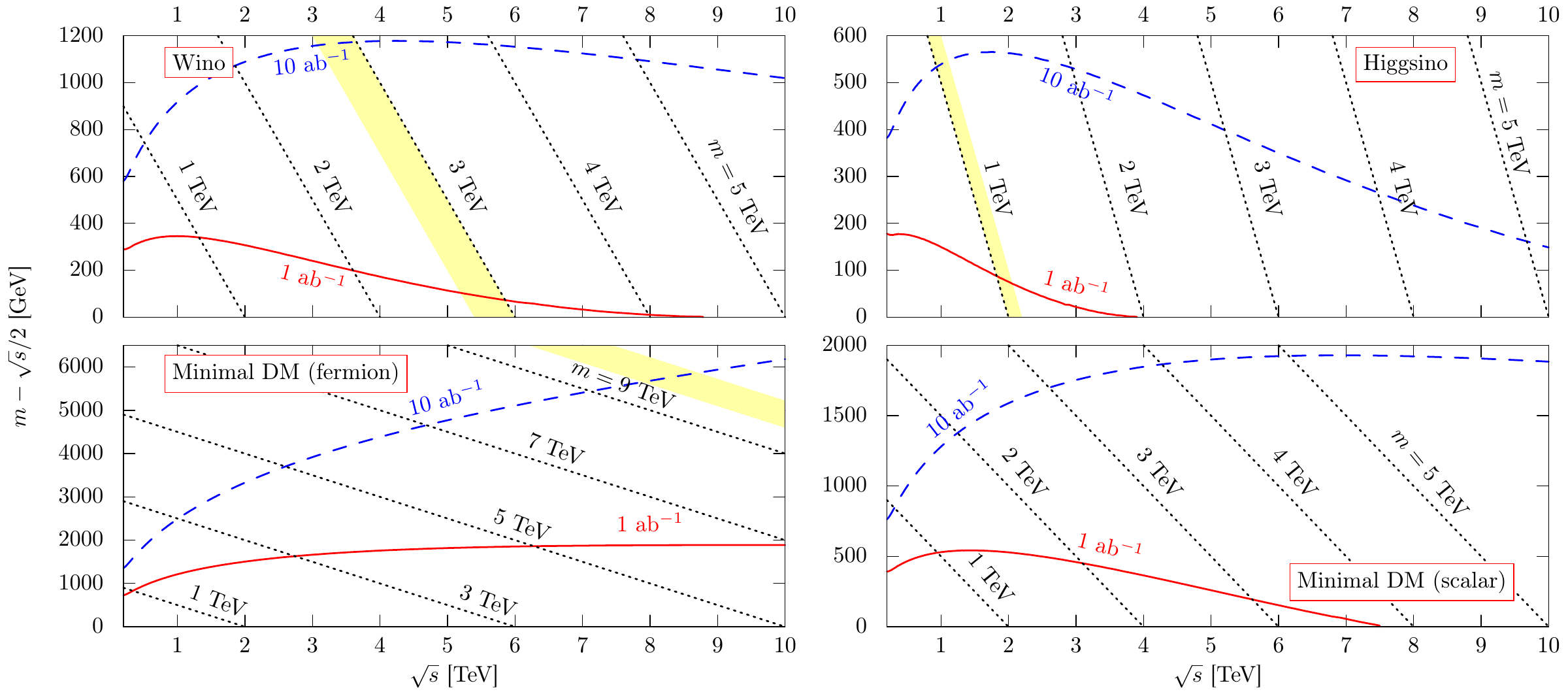}
\caption{\sl \small Ultimate potential of future lepton colliders with large center of mass energy to probe the wino LSP, the Higgsino LSP and the two MDMs discussed in previous subsection. The integrated luminosity is fixed to be $1$ ab$^{-1}$ (red solid lines) and 10 ab$^{-1}$ (blue dashed lines), while the polarizations electron and positron beams are $-80\%$ and $60\%$. Only the statistical uncertainty is taken into account in the analysis. In the yellow-shaded region, the thermal relic abundance of EWIMPs explains the observed abundance of dark matter in the present universe $\Omega h^2 \simeq 0.12$ \cite{Hisano:2006nn, Cirelli:2007xd}}.
\label{fig: TeV_prospect}
\end{figure}

The capability of future lepton colliders to probe the LSP dark matters and the MDM discussed in previous subsection is shown in Fig.\,\ref{fig: TeV_prospect}. The integrated luminosity is fixed to be $L = $1 ab$^{-1}$ (red solid lines) and 10 ab$^{-1}$ (blue dashed lines), while the polarizations electron and positron beams are $-80\%$ and $60\%$. The yellow shaded band represents the region that the observed dark matter density in the universe is explained only by thermal dark matter relics. As we expected in section \ref{subsec: analysis}, the sensitivity reaches are in good agreement with the scaling law $\sim  s^{1/4} L^{1/4}$ although small deviations appear due to logarithm corrections. As can be seen from the figure, the region can be probed through the di-fermion processes when $\sqrt{s}$ is about the dark matter mass required by the WIMP miracle, except the case of the scalar MDM.

\section{Summary}
\label{sec: summary}

In this article, we have studied the capability of future lepton colliders, such as the ILC, CLIC and FCC-ee, to probe EWIMPs indirectly. We have shown that di-fermion production processes $e^- e^+ \to f \bar{f}$ are suitable for this purpose when the mass of an EWIMP is much larger than the beam energy. We have found that the mass larger than the beam energy by 100-1000 GeV is actually detectable when systematic errors to measure the cross sections of the processes are well under control at $O(0.1)\%$ level. We adopt somehow optimistic and simplified assumptions on the collider setup. The systematic errors have actually many origins and thus more complicated. A detailed and realistic analysis will be necessary to conclude the capability of future lepton colliders for this indirect probes, while we expect that it does not alter our result so much and thus the di-fermion processes will play an important role to search for the EWIMP at the colliders.

Let us comment on other channels than di-fermion productions. As we have mentioned in Section\,\ref{sec: bsm contributions}, the effect of EWIMPs on the triple gauge couplings (e.g. $e^-e^+ \to Z/\gamma \to W^-W^+$) is not so useful, as far as $m\gg\sqrt{s}$. However, when the mass and the beam energy are close to each other, $m \simeq \sqrt{s}/2$, the description via dimension six operators is no longer valid. Especially, for an EWIMP with smaller $n$ and $Y$ like a Higgsino, the reach of di-fermion channels is not far above beam energy (see Fig.\,\ref{fig: susy prospect}). In such a case, it is not easy to determine which modes, di-fermion or di-boson, is more suitable to search for EWIMPs. It is therefore interesting to study also on the triple gauge boson couplings as a probe of EWIMPs. Another interesting phenomena may appear when $m \simeq \sqrt{s}/2$. In such a case, a nearly on-shell EWIMP bound states will appear as an intermediate state and may affect the production cross section significantly. Detailed analysis on this effect is beyond the scope of this article and we put it as a future work.

\vspace{0.5cm}
\noindent
{\bf Acknowledgments}
\vspace{0.1cm}\\
\noindent
This work is supported by the Grant-in-Aid for Scientific research from the Ministry of Education, Science, Sports, and Culture (MEXT), Japan No. 26104009 and 26287039 (S.M.), as well as by the World Premier International Research Center Initiative (WPI), MEXT, Japan (K.H., K.I. and S.M.). The work of K.H. and K.I. is supported in part by a JSPS Research Fellowships for Young Scientists. A.K. likes to thank Department of Science and Technology, Government of India, and Council for Scientific and Industrial Research, Government of India, for funding through research projects. He also likes to thank Kavli IPMU for hospitality during the initial stages of the collaboration.

\bibliographystyle{aps}
\bibliography{ref}

\end{document}